\documentclass{aa} 
\usepackage{astron,graphicx}

\newcommand{\Msun}{\mbox{M}_\odot}
\newcommand{\Mch}{M_{\rm Ch}}
\newcommand{\Mni}{M_{\rm Ni}}
\newcommand{\Mej}{M_{\rm ej}}

\begin{document}
\thesaurus{04 (08.19.4; 8.19.5 SN 1991bg; 8.19.5 SN 1991T; 8.19.5 SN 1992A;
8.19.5 SN 1993L; 8.19.5 SN 1994D;)}

\title{SN~Ia light curves and radioactive decay\thanks{Based on ESO 
observations collected at ESO-La Silla (Chile)}}

\author{E. Cappellaro\inst{1} \and P. A. Mazzali\inst{2,3} \and
S. Benetti\inst{4} \and I.J. Danziger\inst{3} \and
M. Turatto\inst{4,1} \and M. Della Valle\inst{5} \and
F. Patat\inst{5}}

\institute{Osservatorio Astronomico di Padova,
vicolo dell'Osservatorio 5, 35122 Padova, Italy
\and
National Astronomical Observatory, Mitaka, Tokyo 181, Japan
\and
Osservatorio Astronomico di Trieste, via G.B. Tiepolo 11, I-34131, 
Trieste, Italy
\and
European Southern Observatory, Alonso de Cordova 3107, Vitacura,
Casilla 19001, Santiago 19, Chile
\and
Dipartimento di Astronomia, Universit\`a di Padova,
vicolo dell'Osservatorio 5, 35122 Padova, Italy}

\date{Received ......; accepted .......}
\maketitle
\begin{abstract}

The absolute $V$ light curves of 5 SNe~Ia, selected to represent the known 
range of absolute luminosities at maximum for this class of objects, are 
presented. Comparison of the long term luminosity evolution shows that the 
differences seen at maximum persist, and actually increase with time, 
reinforcing the notion that intrinsic differences do exist among SNe~Ia. 
Since such differences are not accounted for in the standard progenitor 
scenario, it becomes important to derive constraints for the models 
directly from the observations.  In order to investigate the influence of 
the two most important parameters, that is the masses of the synthesized 
radioactive material and of the ejecta, a simple MC light curve model was 
used to simulate the luminosity evolution from the explosion to very late 
epochs ($\sim 1000$ days).  It was found that the observations require a 
range of a factor 10 in the masses of the radioactive material synthesized 
in the explosion ($\Mni= 0.1-1.1 \Msun$,) and a factor 2 in the total 
mass of the ejecta ($\Mej = 0.7-1.4 \Msun$). Differences of a factor 2 in 
$\Mni$ seem to be present even among `normal' SNe~Ia.

Some evidence was also found that the deposition of the positrons from
Co decay varies from object to object, and with time.  
In particular, the latest HST observations of SN~1992A seem to
imply complete trapping of the positrons.

\end{abstract}


\section{Introduction}

For many years type Ia supernovae (SNe~Ia) were considered very reliable 
standard candles. However, more accurate CCD photometry for a rapidly 
growing number of SNe~Ia has shown the existence of a range in the 
properties of this type of SNe. In particular, a correlation has been 
found between the absolute magnitude at maximum and the shape of the light 
curve \cite{hamuy,riess}.  If this relationship can be well calibrated 
and other parameters are found to play a negligible role, the confidence 
in the use of SNe~Ia as distance indicators may be restored.

A more subtle problem concerns the progenitors of SNe~Ia. The standard
scenario for SNe~Ia is that a degenerate white dwarf in a binary
system accretes matter from a close companion, either red giant or
white dwarf (WD), until it reaches the Chandrasekhar mass ($\Mch=1.4
\Msun$) and explodes, leaving no remnant \cite{nomoto,ww:86}.  By
tuning the relative masses and the separation of the stars in the
binary system it is possible to obtain a very slow evolution and thus
to explain the occurrence of SNe~Ia among population II stars and
hence in elliptical galaxies.  If all SNe~Ia events are triggered by
the WD's mass reaching $\Mch$, independently of the initial
conditions, it is natural to expect some homogeneity in the outcome of
the explosions. One should nevertheless keep in mind the fact that
type Ia supernovae occur in all types of galaxies. This theoretical
view reinforced the empirical use of SNe~Ia as standard
candles. Lately, however, with the advent of larger samples, the
observed diversity among SNe~Ia appears to challenge that paradigm.

In particular, in the following we will show how the simple comparison
of the late light curves of different SNe~Ia requires the existence of
a range in both the mass of radioactive material synthesized in the
SN~Ia explosions and the mass of the ejecta.  Such differences can be
seen not only in the well-known cases of peculiar SNe such as the
bright \object{SN 1991T} and the faint
\object{SN 1991bg}, but also within the group that has hitherto 
preserved the definition of `normal' SN~Ia.

\section{Late light curves}

We have compiled the absolute $V$ light curves of five SNe~Ia, 
namely \object{SN 1991T}, \object{SN 1991bg}, \object{SN 1992A}, 
\object{SN 1993L} and \object{SN 1994D}, combining
unpublished observations with data from the literature.

In principle, bolometric light curves should be used for comparison with 
the models.  Unfortunately, such data are available only for very few SNe, 
even in the best case they are limited to the UVOIR spectral region and, 
with the exception of SN~1987A, they do not cover late epochs. However, 
both observations and spectrum synthesis indicate that in SNe~Ia most of 
the deposited radioactive energy is re-radiated in the optical region and 
that, at phases later than 100d, the bolometric correction is constant and 
small, $m_{\rm bol-m_{\rm V}}\le 0.1-0.2$ mag \cite{sunt92a,wheelhof}.  
Therefore, in the following we simply assume that the bolometric correction 
for the $V$ band is zero for all SNe~Ia. Despite this crude approximation, 
this is not the major source of error, the uncertainties in the parent 
galaxies distance moduli and extinctions being in fact much larger.

The observations were retrieved from the archive of the ESO supernova
monitoring programme \cite{kp}, and are presented here for the first time, 
except for SN~1991bg, whose complete light curve has already been published 
in Turatto et al. \cite*{tur91bg}. These data have been supplemented, for 
the early phases, with published photometry from Phillips et al. 
\cite*{phil91t} for SN~1991T, Sunzeff et al. \cite*{sunt92a} for SN~1992A, 
Filippenko et al. \cite*{fil91bg} and Leibundgut et al. \cite*{lei91bg} 
for SN~1991bg and Patat et al. \cite*{patat} for SN1994D.

Additionally, we included the photometry of SN~1992A obtained 
with WFPC2 on HST on Aug 2, 1994, 926 days after maximum.  
These observations, obtained in the framework of the SINS program 
(Kirshner et al. 1993) \nocite{kir92a}, consist of two sequences of four 
exposures through the F555W and the F439W filters, which are similar 
to the $V$ and $B$ bands, respectively. Exposure times for the
individual frames were 900 sec for F555W and 1200 sec for F439W.  
The individual frames, calibrated in the standard pipeline, have been
retrieved from the HST archive, properly aligned, and combined to
eliminate cosmic rays.  In the combined F555W image we found a stellar
object whose offset from the field stars agrees to within 0.2 arcsec in
both coordinates with the offset of SN~1992A as measured on the last
ground-based observations (note that 1.2 arcsec west of the SN is a 23
mag background galaxy which appears as a stellar object on ground-based 
images).  The SN magnitude, as measured by means of aperture photometry 
and calibrated relative to a sequence of local stars whose magnitudes 
were determined from the ground-based observations, is $V=25.9\pm0.3$.  
The SN is not so evident in the combined F439W image which, however, 
is consistent with a $(B-V)$ colour close to 0.

\begin{figure*} 
\resizebox{\hsize}{!}{\includegraphics*[angle=-90]{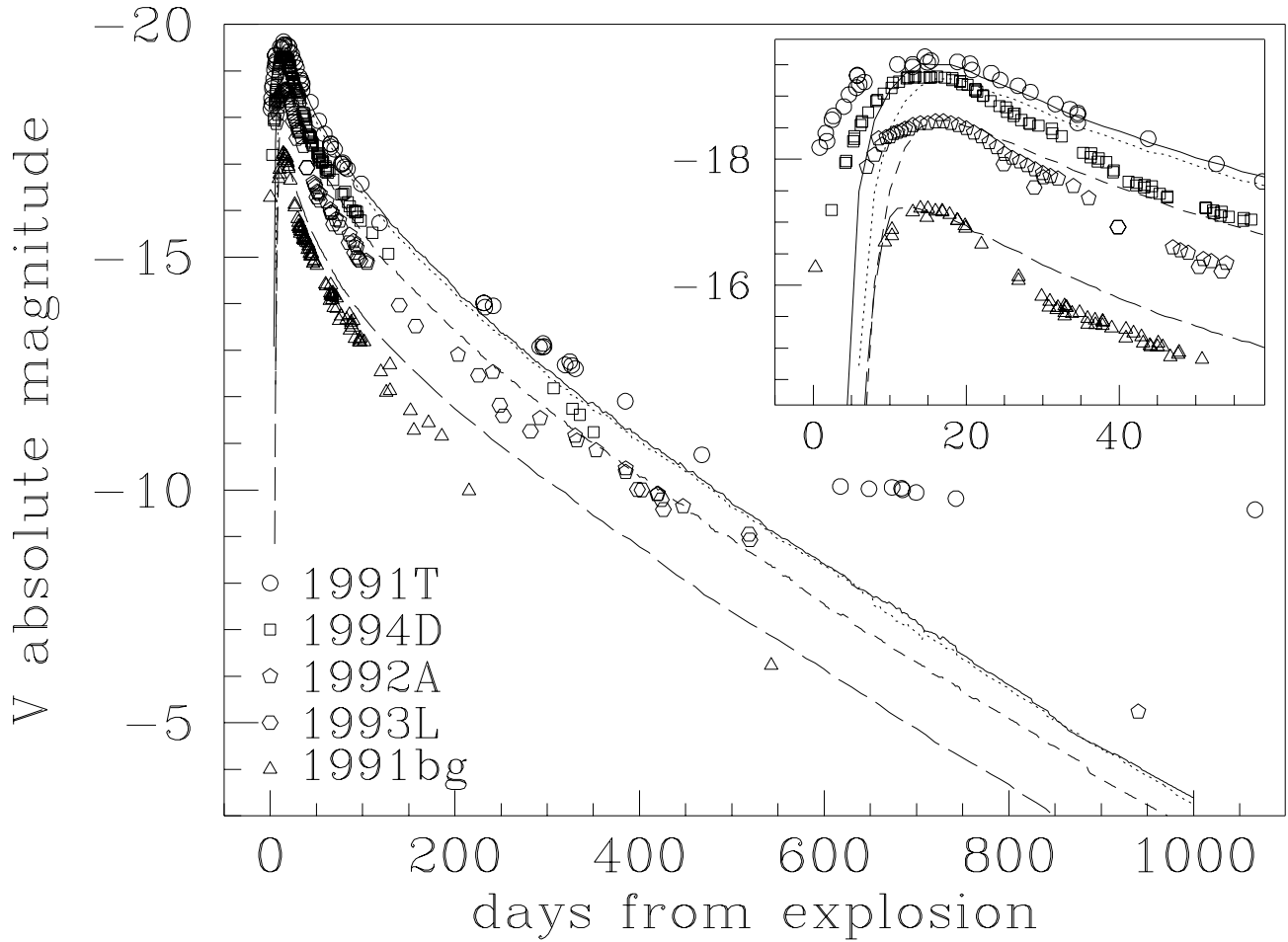}}
\caption{The absolute V light curves of 5 SN~Ia are compared with 4
models with different ejecta and Ni masses. The reference epochs is
the time of explosion estimated from spectrum synthesis and light
curve modelling of the early photospheric phase.  From top to bottom
the models are characterized by $M_{ej}=1.4$, $M_{Ni}=1.1$ (continuous
line); 1.4, 0.8 (dotted); 1.0, 0.4 (short dashed) and 0.7, 0.1 (long
dashed).  For all models $\kappa_e^+ = 7$ is adopted.  In the inset,
the epoch near maximum is shown enlarged.}\label{all}
\end{figure*}

The main data for the five SNe~Ia are listed in Tab.~\ref{tab}, while
the absolute $V$ light curves are shown in Fig.~\ref{all}.  Absolute
magnitudes were computed using the distance moduli given in the Tully
\cite*{tully} catalog, except for SN~1993L, whose parent galaxy is not 
listed there. For this SN we use the distance given in the Leda
extragalactic database\footnote{The Lyon-Meudon Extragalactic Database
(LEDA) is supplied by the LEDA team at the CRAL-Observatoire de Lyon
(France).}. Uncertainty in the absolute magnitudes arises mostly from
the uncertainty concerning the distance scale. (both catalogs we are
using adopt a Hubble constant of 75 km s$^{-1}$ Mpc$^{-1}$).
Although the errors in the absolute magnitudes may be quite
large (of the order of $\pm 1$ mag), the relative distances should be
more reliable (typical errors are $\pm 0.2-0.3$ mag) and so the
differences in magnitude between different SNe are expected to be
real.

Magnitudes have been corrected for the total absorption $A_V$ as 
estimated in the references indicated in Tab.~\ref{tab}.

The reference epochs for the light curves were chosen using the times of 
explosion estimated from spectrum synthesis and light curve modelling of 
the early photospheric phase. These range between 12 and 20 days prior to 
$B$ maximum. The uncertainty in the determination of the time of maximum is 
negligible for our discussion, which is mostly concerned with the very 
late epochs.

\begin{table*}[t]
\caption{SNe Ia data } \label{tab}
\begin{flushleft}
\begin{tabular}{lllcrrllcr}
\hline
 \multicolumn{1}{c}{SN}
 & \multicolumn{1}{c}{Gal}
 & \multicolumn{1}{c}{$V_{\rm max}$}
 & \multicolumn{1}{c}{$(B-V)^*$}
 & \multicolumn{1}{c}{$M-m$}
 & \multicolumn{1}{c}{$A_V$}
 & \multicolumn{1}{c}{$M_V$}
 & \multicolumn{1}{c}{$\Delta m_{15}^B$}
 & \multicolumn{1}{c}{$\Delta m_{300}^V$}
 & \multicolumn{1}{r}{Neb.vel.}
\\
 & & & & & & & & & [km s$^{-1}]$\\
 \hline
 1991bg & NGC4374 & 13.96 &~~0.74 & 31.13 & 0.15$^a$ &$-17.27$ & 1.95$^a$ 
        & 8.4 & 2500 \\
 1993L & IC5270 & 13.2$^{\#}$ &~~0.2~ & 31.52 & 0.23$^{\$}$ &$-18.5^{\#}$ & 
         1.5$^{\#}$  & 7.4 & 8000 \\
 1992A  & NGC1380 & 12.54 &~~0.00 & 31.14 & 0.00$^b$ &$-18.60$ & 1.47$^d$ 
        & 7.3 & 8500 \\
 1994D  & NGC4526 & 11.90 &$-0.08$& 31.13 & 0.08$^c$ &$-19.31$ & 1.26$^e$ 
        & 7.3 & 9500 \\
 1991T  & NGC4527 & 11.50 &~~0.13 & 30.65 & 0.40$^f$ &$-19.55$ & 0.94$^g$ 
        & 6.7 & 10500 \\
\hline
\end{tabular}
\end{flushleft}

$*$  -  measured at the epoch of the B maximum\\
$\$$ -  only galactic absorption included\\
$\#$ -  the SN was discovered ten days after maximum. Its magnitude
at maximum was extrapolated by comparison with SN~1992A.\\
Refs.: $a$ - Turatto et al. \cite*{tur91bg};
$b$ -  Kirshner et al. \cite*{kir92a};
$c$ - Ho \& Filippenko \cite*{hofil};
$d$ - Hamuy et al \cite*{ham:96};
$e$ - Patat et al. \cite*{patat};
$f$ - Mazzali et al. \cite*{maz91t};
$g$ - Phillips \cite*{phil:93}
\end{table*}

In Tab.~\ref{tab} we also list the quantities $\Delta m_{15}^B$, the
difference in $B$ magnitude from maximum to 15d, which is widely used
to characterize the early light curve (col.~8), and $\Delta
m_{300}^V$, the difference in $V$ magnitude from maximum to 300d
(col.~9).  When observations at this epoch were not available, the
value of $\Delta m_{300}^V$ was derived by interpolation between the
closest adjacent observations.  Finally, we report the maximum
expansion velocities of the Fe-nebula (col.~10). These are derived
from models of the emission lines in spectra at epochs around 300d,
except for SN~1991bg, whose spectral evolution was fast and for which
a spectrum at 221d was used (Mazzali et al., 1997).  The velocities
tabulated correspond to the outer velocity of the Fe-nebula for which
synthetic nebular spectra give a best fit to the late-time spectra of
the SNe at hand (Mazzali et al., in preparation).

The SNe have been selected to represent the known range of luminosities 
of SNe~Ia at maximum, going from the faint SN~1991bg to the `average' 
SNe 1992A, 1993L and 1994D and the bright SN~1991T. The full range is 
about 2.3 mag at maximum. In particular, it appears that SN~1994D is 
consistently and significantly brighter than SNe 1992A and 1993L.  
This was also implied by Hamuy et al. \cite*{hamuy} on the basis of the 
different decline rates.  Fig.1 shows that the differences in absolute 
magnitude persist to the very late phases, and actually increase with 
time, mostly reflecting the range in the early decline rates.  
Differences are about 3.5 mag already 2 weeks after maximum, and reach 
about 4 mag at 300d (Tab.~\ref{tab}), after which they apparently stop 
increasing.

The exception is SN~1991T, for which the luminosity after 500d stops
declining. This results from an echo formed as the light emitted by the 
supernova near maximum light was reflected off dust in the CSM or ISM.  
(Schmidt et al., 1994\nocite{schmidt}; Danziger et al. unpublished spectra).  
Since there is no direct relation between the late-time luminosity of 
SN~1991T and the SN remnant itself, in the following we ignore the 
observations of SN~1991T at phases later that 500d.

>From Tab.~\ref{tab} the correlation between the absolute magnitudes at
maximum and the early decline rate ($\Delta m_{15}^B$) is evident.  In
addition, it appears that at late phases \, brighter SNe have larger
Fe-nebula expansion velocities (as already noticed by Danziger, 1994
\nocite{danz94}
and Turatto et al., 1996 \nocite{tur91bg}) and slower luminosity
decline rates.

Another interesting feature shown in Fig.~\ref{all} is that, although 
the late light curves may appear quite linear on short time intervals, 
the decline rate actually slows down in the long run ($>500d$). 
Unfortunately, the evidence is based on only one observation for each of 
the SNe 1991bg, 1992A and, possibly, 1993L. Since observations at such late 
epochs are difficult, and the photometry may be contaminated by phenomena 
not directly related to the remnant itself (eg. echoes in the case of 
SN~1991T or contaminating unresolved stars) this suggestion needs further 
confirmation. Meanwhile, it is worth exploring the implications of 
the effect if real.

\section{Radioactive decay}

Starting with the earliest attempts to model the SN light curves it has 
been clear that the fast expansion of the ejecta would result in a very 
rapid cooling and hence in a very rapid luminosity decline if only the 
thermal energy from the explosion had been available to power the SN.

On the contrary, for almost all types of SNe, the luminosity evolution
is relatively slow.  In particular, at epochs later than 150--200d the
light curves decline almost linearly with rates in the range 0.8--1.5
mag/100d \cite{tur90}, implying the presence of a delayed energy input. 
The linear tail of the light curves corresponds to an exponential 
luminosity decline, suggesting that the radioactive decay of unstable 
heavy elements is a likely energy source, even though the decline rates 
for SNe Ia do not match the decay rates of known radioactive species.

Indeed, model calculations have shown that explosive silicon burning
can produce unstable isotopes of iron group elements, in particular
$^{56}$Ni. $^{56}$Ni decays into $^{56}$Co, emitting $\gamma$-rays 
with an average energy per decay of 1.71 MeV \cite{burro}.  Since the
half-life of $^{56}$Ni is relatively short ($\sim6.1$ days), this decay 
can only be important for the early light curve (the first 1--2 months).

$^{56}$Co is also unstable and decays into $^{56}$Fe. The average
energy available per decay is 3.67 MeV. Although most of this energy 
is released in the form of $\gamma$-rays, a significant fraction of 
the $^{56}$Co decays (19\%) produces positrons. Positrons deposit 
their kinetic energy in the ejecta and then annihilate with electrons, 
producing two photons of energy $E_\gamma = m_e c^2$ each.  The positron 
kinetic energy accounts for about 3.5\% of the total $^{56}$Co decay 
energy \cite{arnett79,axelrod}.
Because of its longer half-life (77.7 days), 
$^{56}$Co decay can power the light curve of SNe for at least 2--3 years.

There is now ample evidence that the radioactive decay of $^{56}$Ni
and $^{56}$Co provides most of the energy during the first 2-3 years,
as was first suggested by Pankey \cite*{pankey} and then,
independently and more quantitatively, by Colgate and McKee
\cite*{colgate}. The most convincing direct evidence has been the
detection of $\gamma$-rays lines from $^{56}$Co decay in the type II
SN~1987A \cite{matz,arnett,palmer} and the temporal behaviour of the
[CoII] lines at $10.52\mu$ \cite{danz} and $1.547\mu$ \cite{varani}.
In the case of many SNe~Ia, modelling of the late-time spectra
indicates the presence of $^{56}$Co lines whose intensity evolution is
in agreement with the prediction from the $^{56}$Co decay
\cite{axelrod,kuchner}.

In the early phases of the SN evolution the density of the ejecta is
still high enough that the $\gamma$-rays from radioactive decay are
trapped in the ejecta and completely thermalized through Compton
scattering with the free electrons. With the expansion the density
decreases and the mean free path of the $\gamma$-rays increases.  
In the case of SNe~II (and of at least a fraction of the SNe~Ib/c) the 
mass of the ejecta is so large ($>3 \Msun$) that even if their density
decreases they remain optically thick to $\gamma$-rays for 2--3 years 
\cite{woosley}. On the other hand, the ejecta become transparent to
optical radiation a few months after maximum, and the observed decline of 
the optical luminosity reflects directly the decline of the radioactive 
energy supply. Indeed, the exponential tails of most SNe~II light curves 
show the same e-folding time as that of the $^{56}$Co energy release.

In the case of SNe~Ia, the ejecta are less massive, while the Ni mass
is larger ($\Mni/\Mej \sim 0.5$ in SNe~Ia vs. $\sim 0.01$ in SNe~II). 
Also, there is evidence that in some cases Ni can be found in the outer 
layers.  This allows an increasing fraction of the $\gamma$-rays to escape 
thermalization as time goes on, so that the light curves of SNe~Ia decline 
more rapidly than the $^{56}$Co energy release. 

\section{A simple light curve model}

The observed differences in the light curves of SNe Ia may be attributed 
to many different factors. Accurate fitting of the light curve of a 
particular SN~Ia may indeed require an ad-hoc explosion model and fine 
tuning of several parameters. However, this approach makes it difficult 
to disentangle the relative importance of the different parameters.  
Since we wanted to isolate the effects of the two arguably most 
interesting physical parameters, $\Mni$ and $\Mej$, we developed a 
simple Monte Carlo code for the calculation of the bolometric light curve 
and did not use any particular explosion model. In particular, we made 
no effort to fit the shape of the light curve near maximum in detail.

Starting from a given density stratification and a distribution of Ni 
within the ejecta, the code computes the deposition of the $\gamma$-rays 
as a function of time with an MC scheme. The $\gamma$-ray energy 
sources are given by Sutherland \& Wheeler (1984), while the opacity 
to $\gamma$-rays is assumed to be gray and to have a constant value 
$\kappa_{\gamma} = 0.027$cm$^2$ g$^{-1}$, in agreement with recent 
calculations \cite{swartz}.  The energy released by the decay of 
$^{56}$Co in the form of positrons (19\% of the total) can only 
be transformed into $\gamma$-rays if the positrons annihilate.

The calculations show that with the expansion the ejecta become rapidly 
transparent to $\gamma$-rays.  In particular, 200 days after the 
explosion the fraction of the $\gamma$-ray energy which is deposited
in a typical model is only $\sim 0.5\%$.  At this time the kinetic
energy of the positrons may become the main contributor to the light
curve, if a major fraction of it is deposited.

The early assumption was that positrons deposit all of their kinetic energy
and annihilate almost on the spot \cite{axelrod}. Subsequently, this idea has
been questioned. For instance, Chan \& Lingenfelter \cite*{chan} suggested
that a significant fraction of the positrons manage to escape, possibly 
because of the presence of radially combed magnetic fields. On the other 
hand, Swartz et al \cite*{swartz} argued that positrons cannot escape at all, 
otherwise a significant fraction of the radioactive decay energy would be lost.

Recently, it has been argued that to fit the late light curve of SNe~Ia 
requires that the ejecta become progressively transparent also to positrons 
\cite{colg:97}, although with a longer time scale than for $\gamma$-rays 
(a typically adopted value is $\kappa_{e^+}=7$ cm$^2$ g$^{-1}$).
Describing the positron deposition by means of a positron opacity appears 
to be a natural approach, and we have adopted it in our light curve code. 
This gives us the freedom to manipulate the positron deposition as well 
as the transport of optical photons. 

Our $\gamma$-ray source function for $^{56}$Co therefore takes the form
\begin{equation}
	S_{Co} = 0.81 s + 0.19 s D_{e^+}
\end{equation}
where 
$ s = 6.78\;10^9 [\exp(-t/\tau_{Co}) - \exp(-t/\tau_{Ni})]$ 
erg~g$^{-1}$~s$^{-1}$
is the rate of energy production from the $^{56}$Co decay (excluding the 
KE of the positrons), $\tau_{Co}$ and $\tau_{Ni}$ are the mean lifetimes 
of $^{56}$Ni and $^{56}$Co, respectively, and $D_{e^+}$ is the positron 
deposition function. 
The positron KE accounts for an additional $\gamma$-ray source, which is 
given by
\begin{equation}
	S_{e^+} = 0.036 s D_{e^+} .
\end{equation}

When the $\gamma$-rays and the positrons deposit their energy, it is
assumed that after thermalization this energy emerges as optical photons, 
so the contribution to $PdV$ work is negligible. These optical photons 
can only contribute to the bolometric luminosity when they manage to 
escape from the ejecta. Given the high ejecta density, especially 
early-on, escape is not instantaneous. This is the reason why SNe~Ia 
reach their $L_{Bol}$ peak only 2--3 weeks after the explosion.  
The random walk of the optical photons in the expanding ejecta is also 
followed with an MC scheme. A gray optical opacity $\kappa_{opt}$ is 
assumed where only a pure scattering opacity obtains.  The ejecta are 
assumed to be purely scattering. Time delay is accounted for by computing 
the time elapsed between successive scatterings, and updating the density 
before a new free flight is started. So finally for each optical energy 
packet we have a `production time', corresponding to the time of emission, 
i.e. the time of deposition of the $\gamma$-ray packet from which the 
optical packet originates, and an `emission time', the time when the packet
eventually escapes from the ejecta. The difference between these two 
times becomes smaller and smaller as time goes on. The code allows us to 
reproduce not only the declining part of the light curve but also, at least 
roughly, the rising branch and the near-maximum phase. For `average' SNe~Ia 
we find that $\kappa_{opt}=0.15$ cm$^2$ g$^{-1}$ gives a reasonable fit 
to the rising branch and to the maximum of the light curve, whereas the 
late light curve is quite insensitive to the value of this parameter.

With these assumptions, the most important parameters for the light
curve modelling are the mass of synthesized $^{56}$Ni ($\Mni$), which
determines the absolute luminosity, and the total mass of the ejecta
($\Mej$), which determines the optical depth for the radiation from the 
radioactive decay. $\kappa_{opt}$ affects mostly the time of maximum and 
its brightness, while $\kappa_{e^+}$ influences the late-time behaviour. 
For larger $\kappa_{opt}$ the maximum occurs later, and it is fainter and 
broader. For smaller $\kappa_{e^+}$ the decay at late times is faster.  

In the past, the consensus was that $\Mni$ and $\Mej$ were exactly 
the same in all SNe~Ia events, making them ideal standard candles. 
Estimates for $\Mni$ were in the range 0.5 to 0.7 $\Msun$, while  
$\Mej=\Mch$. However, there is increasing evidence that both $\Mni$ 
and $\Mej$ can be very different in different objects, although most 
SNe~Ia appear to cluster around the historical values.

Since the SNe we have listed in Table 1 span a range of absolute magnitudes 
at maximum and of decline rates, it is quite likely that they cover a 
range of $\Mni$, and possibly also of $\Mej$. To investigate this question 
further, we used our MC light curve code. As we discussed above, $\Mni$ and 
$\Mej$ can be easily changed as input to the code. We took the approach of 
using a simple code and of not using any particular explosion model for any 
particular object, because literally hundreds of models, involving different 
$\Mej$ and explosion mechanisms, and producing different amounts of $^{56}$Ni, 
have now been produced by various groups, making it difficult to distinguish 
between them on the basis of their ability to fit observed light curves. 

We adopted a W7 density structure, and rescaled it according to the epoch 
and to the values of $\Mni$ and $\Mej$.  We assumed that the Ni resides at 
the centre of the ejecta, in rough agreement with models of SN~Ia explosions. 
The only exception is the model for SN~1991T. In this case, analysis of 
the early-time spectra suggests that there is an outer Ni shell 
(Mazzali et al. 1995). 

Homologous expansion is assumed for rescaling to the appropriate epoch. 
Since different values of $\Mni$ and $\Mej$ should lead to different 
kinetic energies of the ejecta, the $\rho(v)$ dependence is also rescaled. 
If we define the ratios $X_{Ni} = \Mni/0.60\Msun$ and $X_{ej} = \Mej/\Mch$, 
we can rescale the kinetic energy per unit mass according to 
$~KE \propto X_{Ni}/X_{ej}$. 

In doing this we assume that all the KE is produced by burning to $^{56}$Ni, 
neglecting the contribution of incomplete burning to intermediate mass 
elements (IME) such as Si. This is a reasonable approximation as long as 
$M_{IME}/\Mni$ is similar for all values of $\Mni$, which is of course 
not necessarily true. 
In particular, the energy will depend not just on $\Mni$, but also on a 
third parameter, the total mass of the newly synthesized elements. Indeed, in
some cases, like SN~1991bg, the mass of these elements (e.g. Si-Ca, $^{54}Fe$ 
and $^{58}Ni$) is comparable to, and probably greater than $\Mni$. Thus we 
expect our models to be an increasingly poor representation of the real 
properties of the SN the further $\Mni$ departs from the reference W7 value 
of 0.6$\Msun$. 

On the other hand, most SN~Ia explosion models, including models for 
sub-Chandrasekhar explosions (Woosley \& Weaver 1994), produce $\rho(v)$ 
functions whose shape is similar to that of W7, so in this respect our 
approximation should be reasonable. The velocity of a given shell in the 
ejecta is then rescaled according to $v/v_{W7} = (X_{Ni}/X_{ej})^{1/2}$, 
while the density of that shell is given by \,
$\rho / \rho_{W7}  =  X_{ej}^{5/2}  X_{Ni}^{-3/2}$. 

Note that models with a ratio $\Mni/\Mej \simeq 0.43$ have the same KE per 
unit mass as W7 (for which $\Mni/\Mej = 0.60/1.38$), and their density is 
simply the W7 density structure times $X_{ej}$, while models for which 
$X_{ej}^{5/2} = X_{Ni}^{3/2}$ have the same density as W7, but with each 
shell shifted to a velocity $v = v_{W7} (\frac {X_{Ni}}{X_{ej}})^{1/2}$. 
Since in the photospheric epoch the apparent photosphere forms at an 
almost constant value of the density, this may be a useful independent 
tool to investigate the values of $\Mni$ and $\Mej$.
This approach was adopted for all sub-Chandrasekhar models, 
but not for the SN~1991T ones, since the observed outer distribution of 
$^{56}$Ni clearly must have a different effect on the kinetic energy. 
Thus, the KE for the SN~1991T model is probably underestimated.

Apart from the somewhat arbitrary rescaling of the velocity field, several 
other uncertainties must be kept in mind when comparing the model  
bolometric light curve with the observations:

\begin{enumerate}
\item  Though we argued the $V$ and bolometric light curves are not very
different (cf. Sect.~2), they are not the same.  A proper comparison
would require a full NLTE light curve calculation, with the
appropriate SN model. Such calculations do not exist as yet.
Actually,  even the bolometric light curve would require a more complex
calculation than in our simple code. In particular, the opacities may
change as a function of depth in the ejecta, and with time. 

\item The SN may be clumpy, thus changing the deposition rate. This would
also influence the nebular spectrum (Mazzali et al., in preparation).

\item The positon deposition is not well known. We discuss this
point in the next section.

\item The steady-state assumption that the luminosity equals the instantaneous 
energy input breaks down after about 600 days, when the deposition
time for the primary electrons produced by the $\gamma$-rays becomes
long with respect to the dynamical and ionization time scales of the
nebula \cite{axelrod}. The ionization state may then be far from a
steady state solution.

\item The distance and extinction to the various SNe are uncertain. By
using a single source for the distances, we should at least minimize
the relative errors, but the absolute numbers could change. For
instance a Cepheid-based set of distances would require 
much bigger Ni masses.
\end{enumerate}

\section{Results}

We selected the model parameters in order to fit the magnitude at
maximum and the long-term decline rate of the individual SNe. Although 
less emphasis was given to the detailed fit of the light curves,
we comment on the major discrepancies and the possible reasons.

In a first set of models we kept $\kappa_{e^+}=7$ constant, and tried 
to obtain the best fit by varying $\Mni$ and $\Mej$. In general,
increasing $\Mni$ causes a shift of the light curve to brighter
absolute magnitudes, while increasing $\Mej$ delays the time of
maximum and makes the decline of the light curve less steep.

The model fitting to the $V$ light curves of the five SNe~Ia is shown 
in Fig.~\ref{all}. The models used for the fits are as follows. 

For SN~1991bg, $\Mej=0.7\Msun$, $\Mni=0.1\Msun$. This is in agreement with 
findings from modelling of both the photospheric and the nebular-epoch 
spectra (Mazzali et al. 1997), making SN~1991bg the best-studied 
case of a probable sub-Chandrasekhar mass SN~Ia. 

The $V$ light curves of SNe~1992A and 1993L are practically 
indistinguishable once they have been corrected for extinction and
relative distance.  With the assumed distances, they reach a maximum
of about $-18.6$ mag.  The curves are fitted by a model with 
$\Mej=1.0\Msun$, $\Mni=0.4\Msun$, which suggests they may also be
sub-Chandrasekhar events.  The model has the same $\Mni/\Mej$ ratio as W7.

SN~1994D appears to be brighter than SNe~1992A and 1993L. The peak
magnitude ($-19.4$) is well fitted by a model with $\Mej=1.4\Msun$,
$\Mni=0.8\Msun$, implying that the Ni production may be higher than in
`classical' SNe~Ia.  That SN1992A is fainter than average has also
been demonstrated recently by Della Valle et al. (in preparation) who
determined the distance to its parent galaxy using the method of the
globular cluster luminosity function.  

As can be seen from the insert
of Fig.~\ref{all}, a common feature of the models is that the rise to
maximum is steeper and the early decline slower than observed. This
could be improved by adopting different Ni distributions. In general,
placing Ni further out than in the centre of the ejecta makes the
light curve faster.  We have computed a model using the W7 Ni
distribution, which has a `hole' in the central, highest density
regions. This leads to faster escape of the optical photons after
maximum, and produces a faster-declining light curve, which compares
well with the observed ones. Still, W7 seems to be too faint to
reproduce the light curve of SN 1994D and too bright for that of SN
1992A.

It should be noted here that the $V$ light curves of SNe~1994D, 1992A 
and 1993L are very similar in shape. Thus, if the relative distances and 
extinctions were appropriately adjusted, all three SNe could be explained 
with a single explosion model. On the other hand, significant observational 
evidence also exists in favour of there being a difference between SNe~1994D 
and 1992A: SN~1994D is about 0.1 mag bluer at maximum, has a lower SiII line 
velocity near maximum (Patat et al. 1996, Fig.10a, where SNe~1994D, 1989B and 
1990N appear to form one group and SNe~1992A and 1981B another), and broader 
nebular lines, with the SNe just mentioned falling again into two different 
groupings (Mazzali et al, in prep.). Thus, intrinsic differences between 
SNe generally labelled as `classical' Ia are only beginning to receive the 
attention they deserve. 

The last object we tried to fit is SN~1991T. For this SN, analyses of the 
early- (Mazzali et al. 1995) and late-time (Spyromilio et al. 1992) spectra 
suggest that the Ni mass is about 1$\Msun$, and that a significant 
fraction of it is located in the high velocity outer part of the ejecta. 
If we assume $\Mej=\Mch$, a good fit can be obtained for $\Mni=1.1\Msun$, 
of which 0.6$\Msun$ is in the centre and 0.5$\Msun$ is in the outer 
layers, which confirms previous findings. 

Thus, it appears that a range of almost one order of magnitude in $\Mni$ 
is required to fit all the objects, and of at least a factor 2 in $\Mej$. 
The range in $\Mni$ reflects the range in absolute magnitudes at maximum 
rather closely. 

\begin{figure}
\resizebox{\hsize}{!}{\includegraphics*{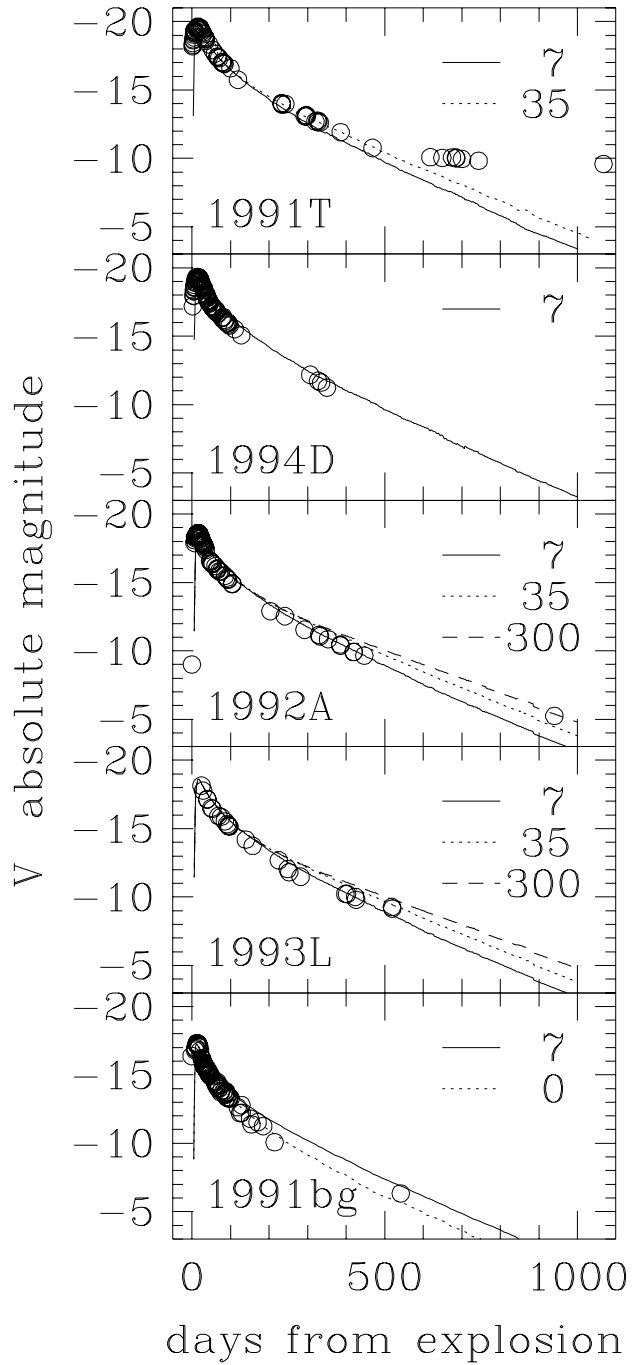}}
\caption{Comparison of the $V$ light curves of different SNe~Ia with
 models for different values of the positron opacity, $\kappa_e^+$. The
 models for the individual SNe are characterized by different values
 of the radioactive Ni and ejecta masses. From top to bottom these are:
1991T:  $\Mej=1.4\Msun$, $\Mni=1.1\Msun$; 
1994D:  $\Mej=1.4\Msun$, $\Mni=0.8\Msun$;
1992A and 1993L:  $\Mej=1.0\Msun$, $\Mni=0.4\Msun$;
1991bg:  $\Mej=0.7\Msun$, $\Mni=0.1\Msun$.} \label{indiv} 
\end{figure}

The models with $\kappa_{e^+}=7$ fit the observations up to 400-500
days well for all the objects except 1991bg.
In the case of SN~1991bg, starting 100--150 days after explosion the
model is brighter than the observations.  To reconcile the model with
the observations, one could further decrease $\Mej$, but this would
lead to a very early maximum and would also cause problems for the
interpretation of the spectra near maximum \cite{maz91bg}.  
Actually, a good fit to the late light curve of SN~1991bg can be obtained 
assuming that the opacity for positrons is much smaller than $\kappa_{e^+}=7$. 
This is shown in the bottom panel of Fig.~\ref{indiv}, where the model is 
calculated for the extreme case of complete transparency of the ejecta to 
positrons ($\kappa_{e^+}=0$).

The opposite trend may be indicated by the SNe~Ia 1993L and 1992A. 
At phases later than 400 days the observed light curves appear to decline 
at a rate slower than the prediction of the model with $\kappa_{e^+}=7$. 
For SN~1994D, no observations are available at these very late epochs.

Finally, the light curve of SN~1991T is compatible with $\kappa_{e^+} > 7$. 
This may be real, but it may also indicate that the fraction of Ni on 
the outside is less than what we have assumed. 
Note, however, that if we distribute all the $1.1 \Msun$ of Ni in the 
centre and compute a rescaled model with $\kappa_{e^+}=7$, this has a 
larger KE than W7, and therefore a lower density, so it actually 
declines faster than the $\kappa_{e^+}=7$ model shown in Fig.2. 
Another possibility is that $\Mej > \Mch$. 
We computed a model with $\Mej = 1.6 \Msun$, $\Mni = 1.2 \Msun$, 
which produces a light curve with a broad maximum and a slow decline. 
This model fits the observations reasonably well at all epochs, 
including the late phases.  
In the phase 60--150 days the model light curve is brighter than the observed 
one, but this is a feature common to all models shown in this paper. Even 
in this case, however, $\kappa_{e^+} > 7$ cannot be ruled out at 300--400 
days. Clearly, SN~1991T deserves a much more detailed investigation 
than has been attempted here. 
 
In principle we cannot exclude the possibility that, in addition to the 
radioactive decay, something else may contribute to the luminosity.  
In particular, given that spectra are not available at these very late 
epochs, we cannot entirely rule out contributions from echoes by 
circumstellar dust, as has been observed for SN~1991T.  
However, the luminosity decline of SN~1992A from 400 to 926 days 
($0.9\pm0.1$ mag/100d) is very close to the $^{56}$Co decay rate 
(0.98 mag/100d,) and other energy sources do not seem to be required.  
As shown in Fig.\ref{indiv}, this is the decline rate expected in the 
case of essentially complete trapping of the positrons 
($\kappa_{e^+} = 300$) and complete transparency to the $\gamma$-rays.

\section{Conclusions}

The observed $V$ light curves of different SNe Ia have been compared
with models of the bolometric light curve obtained varying the values
of $\Mni$ and $\Mej$.
The distribution of Ni was assumed to be central with the exception of the
models for the bright SN 1991T, where an outer Ni shell was also used. 
The $KE$ of the ejecta was rescaled according to $\Mni$ and $\Mej$ in all 
cases except SN~1991T. Both $\Mni$ and $\Mej$ influence the light curve, 
but while $\Mni$ influences mostly the level of the curve, $\Mej$ affects 
also the time of maximum and the decline rate, so that a simultaneous fit 
to both the maximum and the late light curve, up to when the positrons 
begin to dominate, can only be obtained for a single set of $\Mni$ and 
$\Mej$ values. Changes in the distribution of Ni and in the opacities 
between the various models would of course alter this result. 

The comparison of the (late) light curves of SNe~Ia with models confirms
the suggestions derived from the analysis of observed spectra near maximum 
of the existence of a range in both the Ni and the ejecta mass. 
 
It appears that objects which have usually been regarded as `typical' 
SNe~Ia may actually differ by more than 0.5 mag at maximum. This difference
increases with time, suggesting that both $\Mni$ and $\Mej$ may be 
significantly different in these objects (in particular, SN~1994D appears 
to be brighter and to have a larger $\Mej$ than SNe~1992A and 1993L).
Not surprisingly there appears to be a correlation between $\Mni$ and 
$\Mej$ for the sample of 5 objects discussed here.

In general a reasonable fit to the late light curves can be obtained by 
assuming that the ejecta become progressively transparent also to the 
positrons generated in the Co decay.  However, there are indications that 
the positron opacity is not constant from object to object or with time:
the late light curve of SN~1991bg can be fitted if the assumption of 
complete transparency of the ejecta to the positrons is made, while those 
of SNe 1992A and 1993L after 500 days seem to require complete trapping 
of the positrons.

\begin{acknowledgements}
We are pleased to thank Ken Nomoto and Nikolai Chugai for useful discussions. 
P.A.M. acknowkedges receipt of a Foreign Research Fellowship at N.A.O., 
and is grateful to T. Kajino and the Dept. of Astronomy at the University 
of Tokyo for the hospitality.  
\end{acknowledgements}

\bibliographystyle{astron}
\bibliography{/home/enrico/testi/bibliografia}

\end{document}